\documentstyle[12pt]{article}
\newcommand{\sect}[1]{\setcounter{equation}{0}\section{#1}}
\renewcommand{\theequation}{\thesection.\arabic{equation}} \begin{document}
\def\bq{\begin{equation}}
\def\eq{\end{equation}}
\begin{flushright}
{\sl  August, 1996 }\\
{LPTENS-96/51}\\
{UT-Komaba-96/14}\\
\end{flushright}

\begin{center}
{\large\bf SPECTRAL FORM FACTOR IN A RANDOM MATRIX THEORY }
\end{center}

\begin{center}
{\bf E. Br\'ezin$^{a)}$ and S. Hikami$^{b)}$} \end{center}
\vskip 2mm
\begin{center}{$^{a)}$ Laboratoire de Physique Th\'eorique, Ecole Normale
Sup\'erieure}\\ {24 rue Lhomond 75231, Paris Cedex 05, France{\footnote{
Unit\'e propre du centre national de la Recherche Scientifique, Associ\'ee
\`a l'Ecole Normale Sup\'erieure et \`a l'Universit\'e de Paris-Sud}
} }\\
{$^{b)}$ Department of Pure and Applied Sciences, University of Tokyo}\\
{Meguro-ku, Komaba, Tokyo 153, Japan}\\
\end{center}

\vskip 3mm
\begin{abstract}
In the theory of disordered systems the spectral form factor $S(\tau)$, the
Fourier transform of the two-level correlation function with respect to the
difference of energies, is linear for $\tau<\tau_c$ and constant for
$\tau>\tau_c$. Near zero and near $\tau_c$ its exhibits oscillations which
have been discussed in several recent papers. In the problems of mesoscopic
fluctuations and quantum chaos a comparison is often made with random
matrix theory. It turns out that,  even in the simplest Gaussian unitary
ensemble, these oscilllations have not yet been studied there. For random
matrices, the two-level correlation function $\rho(\lambda_1,\lambda_2)$
exhibits several well-known universal properties  in the large N limit. Its
Fourier transform is linear as a consequence  of the short distance
universality of $\rho(\lambda_1,\lambda_2)$. However the cross-over near
zero and $\tau_c$ requires to study these correlations for finite N.
For this purpose we use an exact contour-integral
representation of the two-level correlation function which allows us to
characterize these cross-over oscillatory properties. The method is also
extended  to the time-dependent case.

\end{abstract}
\newpage
\sect{Introduction}

The universal properties of the correlation functions in random matrix
theory have been discussed abundantly
and applied to various fields such as fluctuations in mesoscopic systems or
quantum chaos.

The Fourier transform of a two-level correlation function is a spectral
form factor, which we denote by $S(\tau)$. For the Gaussian unitary
ensemble, in the large N limit,
this form factor $ S(\tau)$ has a simple linear behavior with
$\tau$ up to a critical value $\tau_c = 2 N$,
beyond which it becomes one \cite{Mehta}. This remarkable behavior is due to
the short distance universality of the two-level correlation function
$\rho(\lambda_1,\lambda_2)$\cite{Dyson}.
In the problem of quantum chaos, it is known that the level statistics of
chaotic systems in a certain energy range, agrees with the result of random
matrix theory, and
the linear behavior of $S(\tau)$ has been derived by the method  of
perturbation of periodic orbits \cite{Berry}.

In this article, we evaluate this spectral form factor within  random
matrix theory, in order to characterize the crossover to the linear
behavior in the large N limit. We will investigate the subdominant term,
which has oscillatory behavior, although it can be neglected in the large N
limit. When we consider the derivative of $S(\tau)$ with respect to $\tau$,
this oscillatory behavior is clearly seen, since it becomes of the same
order as the linear term.
We believe that these oscillatory terms near $\tau = 0$ and $\tau = 2 N$,
although  small, are relevant for the discussions of current interest on
oscillations in  disordered metals or in quantum chaos, in the
non-universal regions \cite{Andreev1,AgamAltAnd,AgamFish}.

For the discussion of the crossover to the universal linear behavior, we
derive an exact expression for  finite N of $S(\tau)$ .
Our analysis is based upon the recent calculation of the two-level
correlation function \cite{BH}, in which Kazakov contour-integral
representation \cite{Kazakov} has been used. This representation
has also been used recently for the
Laguerre ensemble;  it made it easy to characterize
the crossover behavior near the edge and near  zero energy for the density
of state \cite{BHZ1}. We consider here a similar crossover behavior for the
two-level correlation function or the spectral form factor.

We find also, after  averaging
this form factor over the energy,
that the corresponding form factor $<S(\tau)>$ is remarkably related, through
a simple
integration, to the density of state $\rho(\tau)$ in the Laguerre ensemble;
this density is known to posses
 a universal oscillatory behavior near the origin \cite{BHZ1,HZ,NS,Andreev}.

We extend the form factor calculations to the time-dependent case, which is
shown to be equivalent to the two-matrix model.
In this case, the singular behavior at the Heisenberg time is smeared 
out.

We further discuss  matrix models  with an external source for the 
correlation functions. We find new characteristic properties 
of the kernel $K_N(\lambda, \mu)$, and its universal behavior. 
The kernel $K_N(\lambda, \mu)$ has lines of zeros in the real 
$(\lambda,\mu)$ plane. We briefly study the zeros of the kernels in the 
two-matrix model and in the model with the external source.
\vskip 5mm
\sect{ Universal
behavior of the form factor}

The two-level correlation function $\rho^{(2)}(\lambda,\mu)$ for the random
matrix model is defined
by
\begin{equation}{\label{2.1}}
\rho^{(2)}(\lambda,\mu) = < {1\over{N}}{\rm Tr} \delta ( \lambda - M)
{1\over{N}} {\rm Tr} \delta ( \mu - M ) > \end{equation}
where the M is an $N \times N$ random Hermitian matrix, and the bracket
means an averaging with respect to the Gaussian distribution;
\begin{equation}\label{2.2}
P(M) = {1\over{Z}} {\rm exp} ( - {N\over{2}} {\rm Tr} M^2 ) \end{equation}

The connected correlation function $\rho_c^{(2)}(\lambda,\mu)$ is obtained
by subtraction of the disconnected part, which is a product of the density
of states $\rho(\lambda)$ and $\rho(\mu)$. This function has a complicated
expression with strong oscillations, which simplifies only in the short
distance limit in which there are a finite number of levels between
$\lambda$ and $\mu$, i. e. for $N(\lambda - \mu)$ finite in the large N
limit. Introducing the scaling variable, 
\begin{equation}\label{2.3}
x = \pi N ( \lambda - \mu ) \rho ({1\over{2}}\lambda + {1\over{2}} \mu )
\end{equation}
and taking the large N limit with a finite x, one finds\cite{Mehta}
\begin{equation}\label{2.4}
\rho^{(2)}_c(\lambda,\mu) \simeq {1\over{N}} \delta(\lambda - \mu)\rho(\lambda)
- \rho(\lambda)\rho(\mu)
{\sin^2 x\over{x^2}}
\end{equation}

The spectral form factor  $S(\tau)$ is defined by
\begin{equation}\label{2.5}
S(\tau) = \int_{-\infty}^{+\infty}d\omega e^{i \omega \tau}
\rho^{(2)}_c( E , E + {\omega} ) \end{equation}
Using  the large N, small $\omega $ limit,  we have, leaving aside  the
delta-function term in (\ref{2.4}), 
\begin{equation}\label{2.6}
\rho^{(2)}_c( E - {\omega\over{2}}, E + {\omega\over{2}} ) \simeq -
{1\over{ \pi^2 N^2}} {\sin^2 [ \pi N \omega \rho(E) ]\over{\omega^2}}
\end{equation}
Then, the Fourier integral is evaluated easily, since
\begin{equation}\label{2.7}
\int_{-\infty}^{+\infty} e^{i\omega t}
{\sin^2(a \omega)\over{\omega^{2}}}d\omega = \left\{\matrix{ {\pi\over{2}}(
2 a - | t |) & \hskip 5mm , |t | < 2a \cr
0 & \hskip 5mm , |t| > 2 a \cr}\right.
\end{equation}
This leads to
\begin{equation}\label{2.8}
 S(\tau) = \left\{\matrix{ {|\tau|\over{2\pi N^2}} - {\rho(E)\over
{ N}} & {\hskip 5mm}, |\tau|< 2 \pi N \rho(E)\cr
0 & {\hskip 5mm}, 
 |\tau| > 2\pi N \rho(E)\cr}\right. \end{equation}
Adding the  $\delta$-function term of (\ref{2.4}), we find that
$ S(\tau)$ vanishes for $\tau = 0$.
>From this result, we find that if $\tau$ is order of N,
then the integration over $\omega$ is dominated by a range of order $1/N$,
and therefore,
the approximation of $\rho^{(2)}(\lambda,\mu)$ 
by its short distance behavior
(\ref{2.6}) is justified.
However, if $\tau$ is order of one, then we have to deal with an
integration over a range in which  $\omega$ is not small,
and we cannot use anymore the short distance universal behavior for
$\rho^{(2)}(\lambda,\mu)$. Therefore, one expects a universal linear
behavior in the range in which $\tau$ is of order $N$.

We have derived in a previous paper \cite{BH} the oscillating short
distance behavior of (\ref{2.6}) by using a method introduced by Kazakov.
This method gives  exact expressions of the correlation functions for
finite N.
It is very convenient for characterizing   the crossovers in  comparison
with the standard  approach based on  orthogonal polynomials.
It consists of
adding to the probability distribution a matrix source, and this external
source is set to zero at the end of the calculation. (In some cases one is
interested in keeping a finite external source,
 as studied recently in \cite{BH}). We thus modify the probability
distribution of the matrix by a source A, an $N\times N$ Hermitian matrix
with eigenvalues $(a_1, \cdots, a_N)$:
\begin{equation}\label{2.9}
P_A(M) = {1\over{Z_A}}{\rm exp}( - {N\over{2}}{\rm Tr}M^2 - N {\rm Tr} A M )
\end{equation}
We consider the average evolution operator with this modified distribution
\begin{equation}\label{2.10}
U_A(t) = < {1\over{N}}{\rm Tr} e^{i t M} > \end{equation}
The density of state $\rho(\lambda)$ is its Fourier transform:
\begin{equation}\label{2.11}
\rho({\lambda}) = \int_{-\infty}^{+\infty} {dt\over{2 \pi}} e^{- i t
\lambda} U_A(t)
\end{equation}
We first  integrate over the unitary matrix $V$ which diagonalizes M, (we
may assume, without loss of generality, that A is a diagonal matrix). This
is done by the well-known Itzykson-Zuber integral \cite{Itzykson},
\begin{equation}\label{2.12}
\int dV {\rm exp}( {\rm Tr} A V B V^{\dag} ) = {{\rm det}({\rm exp}(a_i
b_j))\over{\Delta(A)\Delta(B)}} \end{equation}
where $\Delta(A)$ is the Van der Monde determinant constructed with the
eigenvalues of A:
\begin{equation}\label{2.13}
\Delta(A) = \prod_{i<j}^N (a_i - a_j)
\end{equation}
We are then led to
\begin{eqnarray}\label{2.14}
U_A(t) =&& {1\over{Z_A \Delta(A)}}{1\over{N}}\sum_{\alpha = 1}^{N} \int
dr_1 \cdots dr_N e^{i t r_\alpha} \Delta(r_1,\cdots,r_N) \nonumber\\
&\times& {\rm exp}( - {N\over{2}}\sum r_i^2 - N \sum a_i r_i ) \end{eqnarray}
After integrating over the $r_i$, we obtain
\begin{equation}\label{2.15}
U_A(t) = {1\over{N}} \sum_{\alpha=1}^{N} \prod_{\gamma\neq \alpha}
({a_\alpha - a_\gamma - {i t \over{N}}\over{a_\alpha - a_\gamma}}) e^{-
{t^2\over{2N}} - i t a_\alpha }
\end{equation}
Instead of summing over N terms, one  can write  a contour-integral in the
complex plane,
\begin{equation}\label{2.16}
U_A(t) = - {1\over{i t}} \oint {du\over{2 \pi i}} \prod_{\gamma = 1}^{N}
({u - a_\gamma - {i t\over{N}}\over{ u - a_\gamma}}) e^{- i t u -
{t^2\over{2 N}} } \end{equation}
We may now, and only at this stage,  let the $a_\gamma$ go to zero; we
obtain 
\begin{equation}\label{2.17}
U_0(t) = - {1\over{i t}}e^{- {t^2\over{2 N}}} \oint {du\over{2 \pi i}} e^{-
i t u } ( 1 - { i t \over{ N u}})^N \end{equation}

Similarly the two-level correlation function,
$\rho^{(2)}(\lambda,\mu)$ is
obtained from the Fourier transform $U_A(t_1,t_2)$, after letting A go to
zero \cite{BH},
\begin{equation}\label{2.18}
\rho^{(2)}(\lambda,\mu) = \int\int {dt_1 dt_2\over{(2 \pi)^2}} e^{- i t_1
\lambda - i t_2 \mu}U_0(t_1,t_2) \end{equation}
where $U_A(t_1,t_2)$ is
\begin{equation}\label{2.19}
U_A(t_1,t_2) = <{1\over{N}}{\rm Tr} e^{i t_1 M} {1\over{N}} {\rm Tr} e^{i
t_2 M} >
\end{equation}
The same procedure leads to
\bq\label{2.20}
U_A^{(2)}(t_1,t_2) = {1\over{N^2}}\sum_{\alpha_1,\alpha_2} \int
\prod dr_i
{\Delta(r)\over{\Delta(A)}}e^{-N\sum ({1\over{2}}r_i^2 + r_i a_i) + i (t_1
r_{\alpha_1} + t_2 r_{\alpha_2})} \eq
By integration over the $r_i$, we obtain, after subtraction of the
disconnected part,
a representation  in terms of an integral over two complex variables
\begin{eqnarray}\label{2.21}
U_0(t_1,t_2) &&= -{1\over{N^2}}
\oint {du dv\over{(2\pi i)^2}} e^{- {t_1^2\over{ 2 N}} - {t_2^2\over{2N}} -
i t_1 u - i t_2 v} ( 1 - { i t_1\over{ N u}})^N( 1 - {i t_2 \over{N v}})^N
\nonumber\\
&&\times {1 \over{(u - v - {i t_1\over{N}})(u - v + {i t_2\over{N}})}}
\end{eqnarray}
where the contours are taken around $u = 0$ and $v = 0$. If we let the
contour include the pole, $v = u - {i t_1 \over{N}}$, it gives precisely the
disconnnected term $U_0(t_1 + t_2)$, whose Fourier transform is the
$\delta$ function part of (\ref{2.4}).

We now write the two-level correlation function as the Fourier transform of
$U_0(\lambda_1,\lambda_2)$. In order to show that it takes a factorized
from, we shift the variables $t_1$ and $t_2$  to
$t_1 \rightarrow t_1 - i u N$ and $t_2 \rightarrow t_2 - i v N$. Then, one
finds \begin{eqnarray}\label{2.22}
\rho_c(\lambda_1,\lambda_2) &=& {1\over{N^2}}\int {dt_1\over{2 \pi}} \oint
{dv\over{2 \pi i}}
({ i t_1\over{N v}})^N
{1 \over{v + {i t_1\over{N}}}}e^{- {N\over{2}} v^2 - {t_1^2\over{2N}} - i
t_1 \lambda_1 - N v \lambda_2} \nonumber\\
&&\times \int {dt_2\over{2 \pi}}\oint {du\over{2 \pi i}} (
{ i t_2\over{N u}})^N
{1\over{u + {i t_2\over{N}}}}e^{- {N\over{2}} u^2 - {t_2^2\over{2N}} - i
t_2 \lambda_2 - N u \lambda_1} \nonumber\\
&& = - {1\over{N^2}}
K_N(\lambda_1,\lambda_2) K_N(\lambda_2,\lambda_1) \end{eqnarray}
We have obtained the integral representation for the kernel $K_N(\lambda, 
\mu)$,
\bq\label{2.22a}
K_N(\lambda,\mu) = - \int_{-\infty}^{\infty} {dt\over{2\pi}} \oint 
{du\over{2\pi i}}(- {it\over{Nu}})^N {1\over{u + {it\over{N}}}} 
e^{- {N\over{2}}u^2 - {1\over{2 N}}t^2 - i t \lambda - N u \mu}
\eq
It may be interesting to note that the same integral expression 
is obtained through the orthogonal polynomial method. In Appendix A, we 
give this derivation.

 The expression of $K_N(\lambda_1,\lambda_2)$ may be simplified further  by
the shift $t_1 \rightarrow t + i v N$, \bq\label{2.23}
K_N(\lambda_1,\lambda_2) = \int {dt\over{2 \pi}} \oint {dv\over{2 \pi i}}
{1\over{i t}}
( 1 - { it \over{ N v }})^N
e^{ - {t^2\over{2N}} - i v t - i t \lambda_1 + N v (\lambda_1 - \lambda_2)}
\eq

We may now find  the short distance behavior of $\rho^{(2)}(\lambda_1,
\lambda_2)$ in the large N limit with a finite value of the variable $y=N
(\lambda_1 - \lambda_2)$.
There are several procedures to obtain the oscillating universal form. One
possibility has been discussed in \cite{BH}.
We follow here another procedure for the
purpose of later use.
If we substitute to $v$,  $v \rightarrow i t v$, we may then perform  the
$t$-integration,
\bq\label{2.24}
K_N(\lambda_1,\lambda_2) = {1\over{2 \pi}}\oint {dv\over{ 2 \pi i}}
\sqrt{{\pi\over{{1\over{2N}} - v}}} e^{ - {(v y - \lambda_1)^2\over{ 2 (
{1\over{N}} - 2 v)}} - {1\over{v}}} ( 1 - {1\over{N v}})^N \eq
In the large N limit, we may neglect $1/N$ terms and exponentiate the term
which is a power of  N. We obtain
\bq\label{2.25}
K_N(\lambda_1,\lambda_2) \simeq -{i\over{2\pi}}\oint {dv\over{2 \pi i}} 
 \sqrt{{\pi\over{v}}} e^{{y^2\over{4}}v - {1 -
{\lambda_1^2\over{4}}\over{v}} + {\lambda_1 y\over{2}}} \eq
We change the contour in the complex plane, and we use the following result,
\begin{eqnarray}\label{2.26}
K_{{1\over{2}}}(z) &=& \int_0^{\infty} e^{- z {\rm cosh x}} {\rm cosh
{x\over{2}}} dx \nonumber\\
&=& \sqrt{{\pi\over{2 z}}} e^{- z}
\end{eqnarray}
Where $K_{{1\over{2}}}(z)$ is a modified Bessel function. Then, we obtain
\bq\label{2.27}
K_N(\lambda_1,\lambda_2) \simeq {1\over{\pi y}} e^{{\lambda_1\over{2}}y}
{\rm sin} ( y \sqrt{ 1 - {\lambda_1^2 \over{4}}} )
\eq
The other term $K_N(\lambda_2,\lambda_1)$  is obtained in similar way. Thus
we get, in the large N, finite y, limit,

\begin{eqnarray}\label{2.28}
\rho(\lambda_1,\lambda_2) &=&
- {e^{-{\lambda_1 - \lambda_2
\over{2}}y}\over{\pi^2 y^2}}{\rm sin}({\sqrt{ 4 - \lambda_1^2}
\over{2}} y){\rm sin}({\sqrt{ 4 - \lambda_2^2} \over{2}} y)
\nonumber\\
&\simeq& - {1\over{\pi^2 y^2}} {\rm sin}^2 ( {\sqrt{4 - \lambda_1^2} \over{2}}y )
\end{eqnarray}

We also derive more precise expression for the kernel 
$K_N(\lambda_1, \lambda_2)$ from (\ref{2.22}).
We have
\begin{eqnarray}
K_N(\lambda_1, \lambda_2) &=& \int {dt\over{2 \pi}} \oint {dv\over{2 \pi i}}
({i t\over{N v}})^N {1 \over{ v + {i t\over{N}}}} e^{- {N\over{2}}
v^2 - {t^2\over{2 N}} - i t \lambda_1 - N v \lambda_2}
\nonumber\\
&=& N \int {dt\over{2 \pi}} \oint {dv\over{2 \pi i}} {i^N\over{ v + i t}} e^{ - N f}
\end{eqnarray}
where $f$ is 
\bq
  f = {v^2\over{2}} + {t^2\over{2}} + i t \lambda_1 + v \lambda_2 
- {\rm ln} t + {\rm ln } v
\eq
The saddle point equations for $f$ become
\begin{eqnarray}
   {\partial f\over{\partial v}} &=& v + \lambda_2 + {1\over{v}} = 0
\nonumber\\
{\partial f\over{\partial t}} &=& t + i \lambda_1 - {1\over{t}} = 0
\end{eqnarray}
The four solutions are obtained: $ v = i e^{i\varphi}$, $- i e^{- i \varphi}$,
$ t = e^{ - i \theta}, - e^{i\theta}$. We define $\lambda_1 = 2 {\rm sin}
\theta$ and $ \lambda_2 = 2 {\rm sin} \varphi$.
The Gaussian fluctuation around the saddle point is evaluated by
\begin{eqnarray}
   {1\over{\sqrt{ {\partial^2 f\over{\partial v^2}}}}} &=&
 {e^{\pm i {i\over{2}} \varphi}\over{\sqrt{2 {\rm cos} \varphi}}}\nonumber\\
{1\over{\sqrt{ {\partial^2 f\over{\partial t^2}}}}} &=&
 {e^{\pm i {i\over{2}} \theta}\over{\sqrt{2 {\rm cos} \theta}}}\nonumber\\
\end{eqnarray}
Adding these four saddle point contributions, we have
\begin{eqnarray}\label{a.4}
K_N(\lambda_1,\lambda_2) &=& {e^{{N\over{2}}( {\rm cos}2\theta - 
{\rm cos} 2 \varphi)}\over{8 \pi N \sqrt{ {\rm cos}\theta {\rm cos}
\varphi}}}[ - {{\rm cos} [N( \theta + \varphi + {{\rm sin}2 \theta\over{
2}} + {{\rm sin} 2\varphi\over{2}})] {\rm cos}{1\over{2}}(\theta + 
\varphi)\over{1 + {\rm cos}(\theta + \varphi)}}]\nonumber\\
&+& {{\rm sin}  [N( \theta - \varphi + {1\over{2}}{\rm sin} 2\theta 
- {1\over{2}} {\rm sin} 2 \varphi)]{\rm sin} {1\over{2}}( \theta - 
\varphi )\over{ 1 - {\rm cos} ( \theta - \varphi )}} ]
\end{eqnarray}
When $\lambda_1 - \lambda_2$ is order of ${1\over{N}}$, we make 
approximations in (\ref{a.4}),
\begin{eqnarray}
 \theta - \varphi + {{\rm sin}2\theta\over{2}} - {{\rm sin}2 \varphi
\over{2}} &\simeq& {\rm sin}(\theta - \varphi) + 
({\rm sin} \theta - {\rm sin} \varphi ) {\rm cos}\theta \nonumber\\
&\simeq&
\pi (\lambda_1 - \lambda_2) \rho (\lambda_1)
\end{eqnarray}
where $\rho(\lambda_1) = {\sqrt{ 4 - \lambda_1^2}\over{2\pi}}$, and 
the denominator of (\ref{a.4}) is approximated as
$1 - {\rm cos}(\theta - \varphi) \simeq {1\over{2}}{\rm sin}^2 (
\theta - \varphi)$. Then, in the large N limit for fixed $N(\lambda_1 
- \lambda_2)$ we get the short range universal form of (\ref{2.28}).
Later we will discuss the generalization of (\ref{a.4}) to the 
time dependent case.

We now consider the form factor $ S(\tau)$, which is defined by \ref{2.5}.
>From the expressions of $K_N(0,\omega)$ and $K_N(\omega,0)$ in (\ref{2.24}), 
we have
\begin{eqnarray}\label{2.30}
&&S(\tau) = {1\over{2 \pi }} \int d\omega \oint {du dv\over{(2 \pi i)^2}}
{e^{i \omega \tau} \over{ \sqrt{ {1\over{N}} - 2 u}\sqrt{{1\over{N}} - 2
v}}} e^{-{N^2 v^2 \omega^2\over{2 ({1\over{N}} - 2 v)}} - {\omega^2 ( N u -
1)^2\over{2 ( {1\over{N}} - 2 u)}}} \nonumber\\
&\times& ( 1- {1\over{N u}})^N (1 - {1\over{N v}})^N \end{eqnarray}

In the large N limit, if $\tau$ is order of N, we may use the previous
expressions for re-deriving the universal short distance behavior in
(\ref{2.28}), and
obtain the linear behavior up to $\tau = 2N$. However, for finite N, this
function is complicated, and we need the study of the oscillating part
based on (\ref{2.30}).

\sect{ Oscillatory
behavior of the form factor}

We first integrate out  $\omega$ in (\ref{2.30}), and by shifting $u
\rightarrow {1\over{N}} u$ and $v \rightarrow {1\over{N}} v$, we obtain
\begin{eqnarray}\label{3.1}
S(\tau) &=& {1\over{ \sqrt{ 2 \pi} N^2}} \oint {du dv\over{ ( 2 \pi i)^2}}
e^{- {\tau^2\over{ 2 N D}}} {1\over{\sqrt{ v^2( 1 - 2 u) + ( u - 1)^2 ( 1 -
2 v)}}}\nonumber\\ &\times& ( 1 - {1 \over{u}})^N (1 - {1\over{v}})^N
\end{eqnarray}
where $D$ is given by
\bq\label{3.2}
D = { v^2\over{ 1 - 2 v}} + {( u - 1)^2\over{ 1 - 2 u}} \eq

A quasi-linear behavior with small oscillations follows from this
expression. It is interesting first to compute this contour integral
(\ref{3.1}) for finite N. In Fig. 1 a) the result is shown for $N=7$ . The
correction to the linear behavior is small, but the derivative of $S(\tau)$
with respect to $\tau$, shown in Fig. 1 b), becomes of the same order as
the constant part, at least near $\tau$ = 0. Returning to the analytic
calculation we can obtain exact expressions for this oscillatory
behavior by a saddle point analysis of (\ref{3.1}). For
this purpose, we scale $\tau$ by $\tau = N \tilde \tau$. Then we have
\bq\label{3.3}
S(\tau) = {1\over{\sqrt{2 \pi} N^2}} \oint {du dv\over{ (2\pi i)^2}}
{1\over{\sqrt{ v^2( 1 - 2 u) + ( u - 1)^2 ( 1 - 2 v)}}} e^{- N f}
\eq
where the exponent $f$ is
\bq\label{3.4}
f = { \tilde \tau^2 \over{ 2 D}} - {\rm ln}( 1 - {1\over{u}}) - {\rm ln}( 1
- {1\over{v}} )
\eq
In the large N limit, we look for the saddle points  of $u$ and $v$ in the
complex plane. They are  obtained by
\bq\label{3.5}
{\partial f\over{\partial u}} = 0, {\hskip 10 mm} {\partial f\over{\partial
v}} = 0
\eq
We get thus the two equations,
\bq\label{3.6}
{(1 - 2 u)^2\over{u^2 ( 1 - u)^2}} = {\tilde \tau^2\over{D^2}}, {\hskip 10
mm} {( 1 - 2v)^2\over{v^2 ( 1 - v)^2}} = {\tilde \tau^2 \over{D^2}}
\eq
As solutions of these equations, we have \bq\label{3.7}
{1 - 2 u\over{u( 1 - u)}} = \pm {1 - 2 v\over{v(1 - v)}} \eq
There are four solutions to these equations, but two of them only are
saddle-points : a) for the(+) case, u = v. b) for the(-) case, $u =
{v\over{ 2 v - 1}}$. Although $v = {1 - u \over{1 - 2 u}}$ for the (+) case
and $v = 1 - u$ for the (-) case are solutions, they are not saddle-points
since D vanishes. The case a) u = v leaves us still with four different
solutions. The first one is \bq\label{3.8}
u = v = {1\over{2}} + {i\over{2}} \sqrt{{2 - \tilde \tau \over{ 2 + \tilde
\tau}}}
\eq
The three other solutions  are obtained by the replacement $i \rightarrow
-i$, and $\tilde \tau \rightarrow - \tilde \tau$. Therefore, it is
sufficient to consider explicitly the first case and make the necessary
replacements at the end for the other solutions.  The quantity $D$ becomes
\bq\label{3.9}
D = {i \tilde \tau \over{\sqrt{ 4 - \tilde \tau^2}}} \eq
For this saddle point, $f$ becomes
\bq\label{3.10}
f = i ( 2 \theta - {\rm sin}(2\theta)) - 2 \pi i \eq
where we have put $\tilde \tau = 2 {\rm cos} \theta$. The fluctuation
around this saddle point is obtained by the consideration of the second
derivatives with respect to $u$ and $v$. They are
\begin{eqnarray}\label{3.11}
{\partial^2 f\over{\partial u^2}} &=& {\partial^2 f\over{ \partial v^2}} =
{ 2 i ( 2 + \tilde \tau)^{{3\over{2}}}\over{ \sqrt{ 2 - \tilde \tau}}} ( {2
\over{\tilde \tau }} - \tilde \tau) \nonumber\\
{\partial^2 f\over{\partial u \partial v}} &=& {4 i\over{\tilde \tau}} {( 2
+ \tilde \tau )^{{3\over{2}}}\over{ ( 2 - \tilde \tau )^{{1\over{2}}}}}
\end{eqnarray}
The Gaussian fluctuations around the saddle point produces the inverse of
the square root of a determinant, which is
\begin{eqnarray}\label{3.12}
det f^{''} &=& ({\partial^2 f\over{\partial u^2}})^2 - ({\partial^2 f\over{
\partial u\partial v}})^2\nonumber\\
&=& 4 ( 2 + \tilde \tau)^4
\end{eqnarray}

Thus we obtain from this result,
\begin{eqnarray}\label{3.13}
S(\tau) &\sim&
{e^{- i N ( 2 \theta - {\rm sin} 2\theta )}\over{\sqrt{ 1 - 2 u} \sqrt{1 -
2 v} \sqrt{ D} \sqrt{{\rm det }f^{''}} N^3}}
\nonumber\\
&\sim& {i\over{2}}
{1\over{\sqrt{ 2 i {\rm sin}2\theta } ( 2 + 2 {\rm cos} \theta )N^3}} e^{-
i N ( 2 \theta - {\rm sin}2\theta)} \end{eqnarray}
We now add the other solutions by making the replacements $i \rightarrow -
i$ and $ \tilde\tau \rightarrow - \tilde\tau$, which corresponds to $\theta
\rightarrow \theta + \pi$. Adding these terms, we have \bq\label{3.14}
S_a(\tau) = {{\rm cos}({\pi\over{4}} - N ( 2 \theta - {\rm sin} 2\theta
))\over{ N^3 \sqrt{ 2 {\rm sin} 2\theta}}} [ {1\over{(2 + 2 {\rm cos}\theta
)}} + {1 \over{( 2- 2{\rm cos})}} ]
\eq
where $\tilde \tau = 2 {\rm cos} \theta$; thus  $\theta = 0$ corresponds to
$\tilde \tau = 2$, while $\theta = {\pi\over{2}}$ corresponds to $\tilde
\tau = 0$.

The case b) is quite similar. We have
\bq\label{3.15}
u = {1\over{2}} + {i\over{2}}\sqrt{{ 2 - \tilde\tau\over {2 +
\tilde\tau}}}{\hskip 10mm}
v = {1\over{2}} - {i\over{2}}\sqrt{{2 + \tilde\tau\over {2 - \tilde \tau}}}
\eq
Using the notation $ \tilde \tau = 2 {\rm cos}\theta$, we get \bq
f = i( 2\theta - {\rm sin}2\theta ) - \pi i \eq
and
\bq
D = {i\tilde \tau\over{\sqrt{ 4 - \tilde \tau^2}}} \eq
The ${\rm det} f^{''}$ becomes
\begin{eqnarray}
det f^{''} &=& ({\partial^2 f\over{\partial u^2}})^2 - ({\partial^2 f\over{
\partial u\partial v}})^2\nonumber\\
&=& 4 (4 - \tilde \tau^2)^2
\end{eqnarray}
Adding the four terms obtained by $\tilde \tau \rightarrow - \tilde \tau$
and $i \rightarrow - i$, we have
\bq\label{3.19}
S_b(\tau) = { {\rm cos}( N ( 2\theta - {\rm sin} 2\theta ) - N \pi -
{\pi\over{4}})\over{2\sqrt{\tilde \tau} N^3 ( 4 - \tilde
\tau^2)^{{3\over{4}}}}}
\eq

From these analysis, we obtain the oscillating part of $S(\tau)$
in the large N limit. It is a sum of $S_a(\tau)$ and $S_b(\tau)$. Noting
that the linear part of $S(\tau)$ in (\ref{2.8}) is order $\tau/N^2 =
\tilde \tau/N$, we find that the oscillating part is a nothing but a
correction of order ${1\over{N}}$.
However, if we take a derivative, it becomes of the same order as the
linear term.
We also find that when $\tilde \tau$ is
close to 2, the coefficient of the oscillating part of $S(\tau)$ becomes
large, as shown in (\ref{3.14}) and (\ref{3.19}), and even diverges
at $\tilde \tau = 2$. Therefore, there should be again a crossover near the
critical $\tau_c = 2N$.
Up to now we have considered a fixed energy $E = 0$. In the next section,
we will take instead an average over E, and see that the expression for the
form factor simplifies.
\vskip 5mm

\sect{Average of the form
factor}

We will now
consider the average of $ S(\tau)$ over E, by  simply integrating over E,
\begin{equation}\label{4.1}
<S(\tau)> = \int_{-\infty}^{+\infty} dE S(\tau) \end{equation}
Remarkably, we find that this $<S(\tau)>$ is given analytically in terms of
known functions.

From (\ref{2.5}) and (\ref{2.18}), $S(\tau)$ is written as \begin{eqnarray}
S(\tau) &=& \int d\omega e^{i \omega \tau} \rho (E,E + \omega) \nonumber\\
&=& \int dt_1 e^{- i t_1 E - i \tau E}
U_0(t_1,\tau)
\end{eqnarray}
Thus the integration over $E$ gives simply, 
\begin{eqnarray}\label{4.3}
<S(\tau)> &&= \int dE \int dt_1 e^{- i (t_1 + \tau) E } U_0(t_1, \tau)\nonumber\\
&&= U_0(- \tau, \tau )
\end{eqnarray}
Then we write the following contour-integral representation for $<S(\tau)>$
from (\ref{3.19}),
\begin{equation}
<S(\tau)> = {1\over{N^2}}e^{- {\tau^2\over{N}}}\oint {du dv\over{(2 \pi
i)^2}} e^{- i \tau ( u - v)} ( 1 - { i \tau\over{ N u}})^N ( 1 + { i
\tau\over{ N v}})^N {1\over{( u - v - {i \tau \over{N}})^2}}
\end{equation}
Replacing $u$ by $\tau$ $u$ and  $v$ by $\tau$ $v$,  
putting $\tau^2 = x$, we have
\begin{equation}\label{4.5}
<S(x)> = {1\over{N^2}}\oint {du dv\over{(2 \pi i)^2}} {e^{- i x ( u - v - {i
\over{N}})}\over{(u - v - {i \over{N}})^2}} ( 1- {i \over{N u}})^N ( 1 + {
i\over{ N v}})^N \end{equation}
Taking two derivatives with  respect to $x$, we obtain a simple factorized
expression,
\begin{eqnarray}\label{4.6}
{d^2 <S(x)>\over{d x^2}} &&= - {e^{- {x\over{N}}}\over{N^2}} \oint {du\over{2
\pi i}} e^{- i x u} (1 - {i \over{N u}})^N \oint {dv\over{2 \pi i}} e^{i x
v} ( 1 + { i \over{N v}})^N \nonumber\\
&&= - {e^{- {x\over{N}}}\over{N^2}}[ {d\over{d x}} L_N({x\over{N}})]^2
\end{eqnarray}
where $L_N(x)$ is a Laguerre polynomial.
Remarkably enough, an identical expression has been found earlier, but
for a completely different ensemble and a different quantity.  Indeed the
expression  (\ref{4.6}) has been found in previous work  on the Laguerre
ensemble of random matrices\cite{BHZ1}, in which it was the derivative
of the density of state. The Laguerre ensemble,  also called the chiral
Gaussian unitary ensemble (chGUE) since eigenvalues appear by pairs of
opposite signs, has thus a
 curious relation to the Gaussian unitary ensemble (GUE):
the form factor $<S(\tau)>$ in the GUE is related, for any finite N, to the
density of state  $\rho(\tau)$ of the chGUE. We have not been able to find
a direct proof  of this exact relation  valid for any finite N,  without
calculating both expressions and verifying that they are identical.

The oscillating behavior of (\ref{4.6}) is similar to that of $S(\tau)$.
The oscillating behavior of the density of state for the Laguerre ensemble
near the origin can be seen in the Fig.2 of \cite{HZ}.
In the large N limit, we know that the oscillations of the density of state
near zero energy become universal,
and are given in terms of  Bessel functions.
In view of the previous correspondence we have now to consider the variable
$\tau$ as an energy (although it is a time in the GUE problem!). Near zero
energy, the density of state of the Laguerre ensemble is given by
\begin{equation}\label{3.39}
\rho(\tau) = \tau [ J_0^2(\tau) + J_1^2(\tau) ] \end{equation}
and it shows an oscillating behavior around one. Consequently,
one
understands that the integral of this density of state is proportional to
$\tau$. This is  why we have obtained a linear behavior for $<S(\tau)>$.
However, since the density of state for N large is a semi-circle, and not a
constant ,
 its integral  is no longer proportional to $\tau$. We have
indeed , in the large N limit,
by  integrating  the semi-circle line
\begin{eqnarray}
<S(\tau)> &=& \int_{0}^{\tau} dx \sqrt{ 1 - ({x\over{2 N}})^2} \nonumber\\
&=&
{\tau\sqrt{ 1 - ({\tau\over{2 N}})^2}\over{4 N}} + {{\rm
arcsin}({\tau\over{2N}})\over{2}}
\end{eqnarray}
Beyond the critical value $\tau = 2N$, it remains equal to one, and it
approaches smoothly this limit. Therefore, taking into account the fact
that  the density of state is not constant, the singularity is smoothed
out.  In fig.3 $<S(\tau)>$ is represented in the large N limit. Near the
energy zero, the oscillatory behavior of $<S(\tau)>$, following from the
Bessel functions of (\ref{3.39}), becomes universal . We have shown in
\cite{BHZ1}, by  the same contour-integral representation, 
that there is a
crossover from the bulk to the zero energy region, which
is described universally by the function (\ref{3.39}). 
 We have also found,
in a model consisting of a lattice of coupled matrices, that this
oscillating behavior is model-independent. Near $\tau = 2N$, the crossover
behavior has been also studied in \cite{BHZ1}. It is given by the square of
an Airy function (see eq. (3.37) of \cite{BHZ1}); this crossover is also
known to be universal.
\vskip 5mm

\sect{ Time-dependent case }

    We now proceed to investigate the time dependent correlation 
function and its Fourier transform, the dynamical form factor.
In the large N limit, the universal form of this time dependent 
correlation function has been discussed
\cite{BZ2,Beenakker,Szafer}. We will consider this problem 
by the contour integral representation, which is valid for finite N,
and evaluate the form factor $S(\tau)$ for a fixed time $t$.
For a finite $t$, we will find that $S(\tau)$ shows different 
behavior 
compared to the previous linear behavior about $\tau$.

We consider the N by N Hermitian matrix $M$, which depends 
upon a time $t$. The  time-dependent  correlation function is 
defined by
\bq\label{5.1}
  \rho (\lambda, \mu ; t) = < {1\over{N}}{\rm Tr}
\delta ( \lambda - M(t_1)) {1\over{N}}{\rm Tr}\delta 
( \mu - M(t_2)) >
\eq
where $t = t_1 - t_2$ and $t_1$ and $t_2$ are different times.
This is written as a Fourier transform of the following 
quantity $U(\alpha,\beta)$,  

\bq\label{5.2}
 U(\alpha,\beta) = {1\over{N^2}} < {\rm Tr} e^{i\alpha M(t_1)}{\rm Tr}e^{i \beta M(t_2)}
 >
\eq
We use the new set of variables $\alpha$ and $\beta$ for 
the Fourier transform variables, instead of $t_1$ and $t_2$.
To avoid the confusion, we use $t_1$ and $t_2$ as time.

We show  exactly that the correlation function (\ref{5.1}) 
reduces to the 
correlation function of the two-matrix model in the Gaussian 
ensemble; the c = 1 problem is described by the two-matrix 
model. This correspondence may be derived by other arguments
\cite{Shastry}.  
 We here follow the path integral 
method, which can show explicitly that this equivalence to 
two-matrix model holds for any finite N.

  By considering the following hamiltonian $H$,
\bq\label{5.3}
   H = {1\over{2}} {\rm Tr} (p^2 + M^2),
\eq
where $p = \dot M$ and $M$ is $N\times N$ Hermitian matrix,
we write $U(\alpha,\beta)$ in (\ref{5.2}) as
\bq\label{5.4}
   U(\alpha,\beta) = {1\over{N^2}}
<0|e^{Ht_1}({\rm Tr} e^{i \alpha M})e^{H(t_2 - t_1)}
   ({\rm Tr}e^{i \beta M})e^{- H t_2}|0>.
\eq
We use the path integral formulation, and we define
\bq\label{5.5}
    <A| e^{- \tilde\beta H}|B> = \int_{M(\tilde\beta)= A,M(0)=B}DMe^{-{1\over{2}}{\rm Tr}
        \int_0^{\tilde\beta} (\dot M^2 + M^2)dt}
\eq
Then  $U(\alpha,\beta)$ is expressed by
\begin{eqnarray}\label{5.6}
  U(\alpha,\beta) &=& {1\over{N^2}}
\int dA dB <0|e^{Ht_1}|A><A|({\rm Tr}e^{i \alpha M}) e^{H(t_2 - t_1)}({\rm tr}e^{i\beta M})|B>\nonumber\\
&&\times <B|e^{-Ht_2}|0>
\end{eqnarray}
Noting that the ground state energy of the 
free independent $N^2$ fermions is  $N^2/2$, we have
\bq\label{5.7}
   <0|e^{Ht_1}|A> = e^{{N^2\over{2}}t_1}e^{-{1\over{2}}
 {\rm Tr} A^2}
\eq
The solution of $\ddot M = M$, becomes
\bq\label{5.8}
M(t) = B {\rm ch} t + {{\rm sh}t\over{{\rm sh}\tilde \beta}}
(A - B {\rm ch}
\tilde\beta)
\eq
Then we are able to write the action in (\ref{5.5}) by the matrices 
$A$ and $B$,
\begin{eqnarray}\label{5.9}
   {1\over{2}}{\rm Tr}\int_0^{\tilde\beta} (\dot M^2 + M^2)dt &=& 
 {1\over{2}}
{\rm Tr}(\dot M M)|_0^{\tilde\beta}\nonumber\\
&=&{1\over{2{\rm sh}\tilde\beta}}[(A^2 + B^2){\rm ch}
\tilde\beta - 2 A B]
\end{eqnarray}
Denoting $\tilde\beta$ by a time $t$,  and taking the fluctuation part, 
we get
\begin{eqnarray}\label{5.10}
  &&U(\alpha,\beta) = {1\over{N^2}}
\left( {e^t\over{{\rm sh}t}}\right)^{{N^2\over{2}}}
\int dA dB ({\rm Tr}e^{i \alpha A})({\rm Tr} e^{i \beta B})
\nonumber\\
&&\times e^{- {1\over{2{\rm sh}t}}{\rm Tr}
[(A^2 + B^2) e^t - 2 A B]}
\end{eqnarray}
Thus the problem reduces exactly to a calculation of the 
correlation function for the two-matrix model,  in which matrices
A and B are linearly  coupled.

The  correlation function for two-matrix model 
has been studied by D'anna, Brezin and Zee\cite{Danna}
 by the orthogonal polynomial method for finite N.
Although we can use their result, it is more convenient 
to use the contour integral representation for the correlation
function.

By making the change of variables of  A, B and $\alpha$, 
$\beta$ by a 
factor $\sqrt{e^{-t}{\rm sh }t} $,  we obtain a simple 
expression for (\ref{5.10}),
\bq\label{5.11}
  U(\alpha,\beta) = {1\over{Z}} \int dA dB {\rm Tr} e^{i \alpha A} {\rm Tr}
e^{i \beta B} e^{-{1\over{2}}{\rm Tr} ( A^2 + B^2 - 2 c A B)}
\eq
where 
$c =  e^{-t}$.  Note that we scaled $\alpha$ and $\beta$ by a 
factor $\sqrt{ e^{-t}{\rm sh} t}$, the variables $\lambda$ and $\mu$ 
of the two-point correlation function should be modified by this 
factor for the mapping of the time-dependent case to the  two-matrix 
model. 
We now go back to the notation, in which 
two matrices are given by $M_1$ and $M_2$. We 
denote the  matrices  A and B in (\ref{5.11}) by
$M_1$ and $M_2$.
We introduce the external matrix 
A, which is coupled to matrix $M_1$.
The Gaussian distribution is given by 
\begin{eqnarray}\label{5.12}
P_A(M_1,M_2) &=& {1\over{Z_A}}
e^{ - H_{1,2}}\nonumber\\
 H_{1,2}&=&{1\over{2}}{\rm Tr} M_1^2 +{1\over{2}}{\rm Tr}  M_2^2 - c {\rm Tr} M_1
 M_2 + {\rm Tr} A M_1
\end{eqnarray}

 The density of state $\rho(\lambda)$ is 
given by the Fourier transform of
\bq\label{5.13}
   U_A(z) = < {1\over{N}} {\rm Tr} e^{i z M_1} >
\eq
The calculation of this $\rho(\lambda)$ is similar to the 
one matrix case. The integration over $M_2$, which has eigenvalues $\xi_i$,
 is performed by the help of  Itzykson-Zuber formula. We denote the 
eigenvalues of $M_1$ by $r_i$. The integration over $\xi$ becomes
\bq\label{5.14}
 \int d\xi\prod_{i<j} (\xi_i - \xi_j) e^{-{N\over{2}}\sum \xi_i^2 - c N \sum \xi_i r_i}
 = \prod_{i<j}  (r_i - r_j)e^{{Nc^2\over{2}}\sum{r_i^2}}
\eq
Then we are left with the  integration about $r_i$,
\bq\label{5.15}
U_A(z) = {1\over{\Delta(A)}}\sum_{\alpha=1}^N
 \int d r \prod_{i<j} (r_i - r_j)
e^{-{N\over{2}}(1 - c^2)\sum_i r_i^2 - N \sum_i
 a_i r_i + i z r_\alpha}
\eq
Therefore, by the contour integration, we have  
by letting $a_i$ goes to zero,
\bq\label{5.16}
  U_0(z) = - {\sqrt{1 - c^2}\over{i t}}\oint {du\over{2 \pi i}}
  ( 1 - {iz\over{Nu\sqrt{1 - c^2}}})^N
e^{- {i z u\over{\sqrt{1 - c^2}}}- {z^2\over
  {2N(1 - c^2)}}}
\eq
We have the same density of state as one matrix case except
the scaling factor $(1 - c^2)$,
\bq\label{5.17}
   \rho(\lambda) = \sqrt{1 - c^2} \rho_0(\sqrt{1 - c^2}\lambda)
\eq
where $\rho_0(\lambda)$ is the density of state for the one matrix model.
In the large N limit, this density of state becomes
\bq\label{5.18}
   \rho(\lambda) = {\sqrt{1 - c^2}\over{2 \pi}} \sqrt{ 4 - ( 1 - c^2)\lambda^2}
\eq
which is normalized to be one by the  integration.

The two-level correlation function is given by 
\bq\label{5.19}
\rho^{(2)}(\lambda,\mu) = \int\int {dz_1 dz_2
\over{(2\pi)^2}} e^{- i z_1 \lambda - i z_2 \mu} U_0(z_1,z_2)
\eq
where $U_0(z_1,z_2)$ is
\bq\label{5.20}
   U_0(z_1,z_2) = < {1\over{N}} {\rm Tr} e^{i z_1 M_1} {1\over{N}} 
{\rm Tr}
   e^{i z_2 M_2} >
\eq
By the integration over the eigenvalues $r_i$ of $M_1$, and $\xi_i$ of $M_2$, and 
keeping the eigenvalues $a_i$ of the external matrix A, we have the following 
expression, 
\begin{eqnarray}\label{5.21}
    &&U_A(z_1,z_2) = {1\over{N^2}}\sum_{\alpha_1,\alpha_2} 
    {\prod_{i<j} (a_i - a_j - {iz_1\over{N}}
    (\delta_{i,\alpha_1} - \delta_{j,\alpha_1}) 
      - {i z_2\over{cN}}
    (\delta_{i,\alpha_2} - \delta_{j,\alpha_2}))
      \over{\prod_{i<j}(a_i - a_j)}}\nonumber\\
    &\times& e^{  - {i z_1\over
{1 - c^2}} a_{\alpha_1} - { i z_2 c \over{1 - c^2}}a_{\alpha_2} - 
     {z_1^2\over{2 N ( 1 - c^2)}} -{ z_2^2\over{2 N ( 1 - c^2)}}
- {c z_1 z_2 \over{ N ( 1 - c^2)}}\delta_{\alpha_1,\alpha_2} }
\end{eqnarray}
The double sum for $\alpha_1$ and $\alpha_2$ is divided 
into two parts. The part $\alpha_1= \alpha_2$ is written by the contour-integral representation:
\begin{eqnarray}\label{5.22}
 U_A^{I}(z_1,z_2) =  &&- {1\over{i N (z_1 + {z_2\over{c}})}}\oint {du\over{2\pi i}}
    [ 1 - {i\over{N u}}(z_1 + {z_2\over{c}})]^N
   \nonumber\\
   &\times& e^{ ( - {iz_1\over{1 - c^2}} - {i z_2 c\over{1 - c^2}})u
   - {z_1^2\over{2N(1-c^2)}} - {z_2^2\over{2N(1 - c^2)}} - {c z_1 
z_2\over
{N(1 - c^2)}} }
\end{eqnarray}
The Fourier transform of this quantity
 becomes by the change of variable $u$ to
 $u (z_1 + {z_2\over{c}})$,
\begin{eqnarray}\label{5.23}
  \rho^{I}(\lambda,\mu) &=& {i\over{N}}
 \int {dz_1 dz_2\over{(2\pi)^2}}
 \oint {du\over{2 \pi i}} ( 1 - {i\over{Nu}})^N 
e^{- i ( z_1 + z_2 c)(z_1 + {z_2\over{c}}){u\over{ 1 - c^2}}}
\nonumber\\
 &\times & e^{- {z_1^2 + z_2^2 + 2 c z_1 z_2\over{2N(1 - c^2)}} - 
i z_1 \lambda - i z_2 \mu }
\end{eqnarray}
This part is further simplified by the change of variables
$z_1$ to ${1\over{\sqrt{ 1 - c^2}}} ( z_1 - c z_2 )$ and 
$z_2$ to ${1\over{\sqrt{ 1 - c^2 }}} ( z_2 - c z_1 )$.
By integration over $z_1$, we have
\begin{eqnarray}\label{5.24}
  \rho^{I}(\lambda,\mu)  &=& {i\over{N}} {1\over{( 1 - c^2)^2}}
e^{-{N\over{2}} ( \lambda - c \mu )^2}
\int {dz\over{2 \pi}} \oint {du\over{2 \pi i}}
( 1 - {i \over{N u}} )^N e^{- i \mu z + N u z ( \mu - {\lambda \over
{c}} ) }\nonumber\\
&\times& e^{- {i\over{1 - c^2}} u z^2 - {N\over{2 c^2}} u^2 z^2 
- {1\over{2 N ( 1 - c^2)}} z^2 } 
\end{eqnarray}

The remaining $\alpha_1 \neq \alpha_2$ part is given
  after letting $a_\gamma$ go to zero,
\begin{eqnarray}\label{5.25}
    && U_0(z_1,z_2) = - { c \over{z_1 z_2}}
\oint {du dv\over{(2\pi i)^2}} ( 1 - {i z_1 \over{Nu}})^N
( 1 - {i z_2 \over{c N v}})^N
 \nonumber\\
&\times& [ 1 - {z_1 z_2\over{c N^2 ( u - v - {i z_1\over{N}})
( u - v + {i z_2
\over{c N}})}} ] e^{ - {i z_1 u \over{1 - c^2}} - {i z_2 c v \over 
{1 - c^2}} - {z_1^2\over {2 N ( 1-c^2)}}
 - {z_2^2\over{2 N ( 1 - c^2)}} }\nonumber\\
\end{eqnarray}
 This expression includes both a disconnected part and 
a connected part.
The disconnected part has a factorized form and it corresponds to 
the first term in the bracket.
This term, indeed by shifting $v \rightarrow v/c$, becomes the 
product of the density of states $\rho(\lambda)$ and $\rho(
\mu)$. 
Therefore, after subtracting this disconnected part, we obtain the connected 
part:
\begin{eqnarray}\label{5.26}
U_0(z_1,z_2) &=& - {1\over{N^2}}
\oint {du dv\over{(2 \pi i)^2}} ( 1 - { i z_1 \over
{N u}})^N ( 1 - { i z_2\over{c N v}})^N 
{1\over{(u - v - { i z_1\over{N}})
( u - v + { i z_2\over{ c N}})}}\nonumber\\
 &\times& e^{ - {i z_1 u\over{1 - c^2}} - { i z_2 v c\over{ 1 - c^2}}
- { z_1^2\over{ 2 N ( 1 - c^2)}} - {z_2^2\over{2 N(1 - c^2)}} }
\end{eqnarray}
where the contour-integrals are taken around $u = 0$ and $v = 0$.
If we include the contour-integration  around the pole $v = u - {i z_1
\over{N}}$, we obtain precisely the same term as (\ref{5.22}).
Therefore we use this representation for the whole expression, 
including the term of (\ref{5.21}), by taking the contour around 
both $v = 0$ and $ v = u - iz_1/N$.

The expression for the two-matrix connected correlation 
function $\rho_c^{(2)}(\lambda,\mu)$, which is obtained by the 
Fourier transform of $U_0(z_1,z_2)$, has a factorized form when 
we consider the contribution from $u = 0$ and $ v = 0$. 
If $z_1$ and $z_2$ are replaced by $z_1 = z_1 - i u N$ 
and $z_2 = z_2 - i v c N$, we have a multiplicative form,
\begin{eqnarray}\label{5.27}
&&\rho_c^{(2)}(\lambda,\mu) = - \oint {du\over{2 \pi i}}
\int {dz_2\over{2 \pi}}
({z_2\over{c u N}})^N {1\over{u + {i z_2\over{cN}}}}
 e^{-{N u^2\over{2(1 - c^2)}} -{ z_2^2\over{2 N ( 1 - c^2)}} - i z_2 
\mu - u N \lambda}\nonumber\\
&&\times \oint {dv\over{2\pi i}}\int {dz_1\over{2\pi}}
 ({z_1\over{v N}})^N {1\over{v + {i z_1\over{N}}}}
e^{- {v^2 c^2 N\over{2(1-c^2)}} - {z_1^2\over{2 N (1 - c^2)}} - i z_1 
\lambda - v c \mu N}
\end{eqnarray}

We write this expression as 
\bq\label{5.28}
  \rho^{II}(\lambda,\mu) = - K_N(\lambda, \mu) \bar K_N(
\lambda,\mu)
\eq
We find the previous first part $\rho^{I}(\lambda, \mu)$  
of (\ref{5.24}) is also 
expressed by
\bq\label{5.29}
   \rho^{I}(\lambda, \mu) ={1\over{N}}
 K_N(\lambda,\mu) e^{-{N\over{2}}
(\lambda - c \mu )^2}
\eq
where $K_N(\lambda, \mu) $ is given by
\begin{eqnarray}\label{5.30}
  K_N(\lambda, \mu) &=& i^N ( - i N) \oint {du\over{2 \pi i}}
\int {dz\over{2 \pi}} ( 1 - {i\over{u N}})^N e^{- {Nu^2 z^2\over{2 c^2}}
- {i u z^2 \over{1 - c^2}} + z u \mu N - i z \mu }\nonumber\\
&\times& e^{- {N \lambda u z\over{c}} - {z^2\over{ 2 N ( 1 - c^2)}}}
\end{eqnarray}

This kernel $K_N(\lambda, \mu)$ is written as a sum of  Hermite 
polynomials  $H_n(x)$, 
\bq\label{5.31}
 K_N(\lambda, \mu) = e^{- {N\over{2}} ( 1 - c^2) \mu^2 }
{1\over{N}} \sum_{n=0}^{N-1} {1\over{c^n}} {H_n( \beta \lambda) 
H_n(\beta \mu)\over{n!}}
\eq
where $\beta$ = $\sqrt{{N\over{2}}( 1 - c^2)}$.
The other kernel $\bar K_N(\lambda, \mu)$ is given
\bq\label{5.32}
  \bar K_N(\lambda, \mu) = {1\over{N}}e^{-{N\over{2}}( 1- c^2)
\lambda^2} \sum_{n=0}^{N-1}
c^n{ H_n(\beta \lambda) H_n(\beta \mu)\over{n!}}
\eq
It follows from these expressions that $\rho^{I}(\lambda, \mu)$ and $\rho^{II}(\lambda, 
\mu)$ are invariant under exchange of $\lambda$ and $\mu$.
The expressions of (\ref{5.28}) and (\ref{5.29}) for the correlation function by the kernel $K_N(\lambda, \mu)$ agrees with the result 
obtained by the method of orthogonal polynomials \cite{Danna}. The 
difference is only the exponential Gaussian factor in $K_N(\lambda,
\mu)$ and $\bar K_N(\lambda, \mu)$, and this difference disappears 
for the product of these two kernel.

In the large N limit, we expect to recover the usual universality. 
We return to  the expression of (\ref{5.26})
 and  neglect the terms ${i z_1\over{N}}$
 and ${i z_2\over{cN}}$ in the denominator.
We then  exponentiate the  powers of N.   
 The integrals over $z_1$ and $z_2$ are Gaussian, and it 
leads to
\begin{eqnarray}\label{5.33}
  \rho^{II}(\lambda, \mu) &=& {(1 - c^2)\over{2 \pi N}}
\oint {du dv\over{(2 \pi i)^2}}{1 \over{ ( u - v )^2}}
e^{-{N\over{2( 1 - c^2)}}( u + {1 - c^2\over{u}} + 
( 1 - c^2 ) \lambda )^2}
\nonumber\\
&\times& e^{-{N\over{2 ( 1 - c^2 )}}( v c + { 1 - c^2\over{ c v}} + 
( 1 - c^2 ) \mu )^2 }\nonumber\\
&=& {(1 - c^2)\over{2 \pi N}}
\oint {du dv\over{(2 \pi i)^2}}{1 \over{ ( u - v )^2}} e^{-N(f_1(u) 
+ f_2(v))}
\end{eqnarray}

We use the saddle points of $u(\lambda)$ and $u(\mu)$, 
which are the solutions of ${\partial f_1\over{\partial u}} = 0$
and ${\partial f_2 \over{\partial v}} = 0$, i. e. 
\begin{eqnarray}\label{5.34}
  &&u^2 + \lambda (1 - c^2) u + ( 1 - c^2) = 0\nonumber\\
  &&v^2 + {1 - c^2 \over{c}} \mu v + {1 - c^2 \over{ c^2}} = 0
\end{eqnarray}

Taking into account the fluctuations around the saddle point, 
we have
\bq\label{5.35}
   \rho^{II}(\lambda,\mu) = - {1 - c^2\over{ 2 \pi^2 N^2}} \sum
{1\over{(u(\lambda) - v(\mu))^2}} 
{1\over{\sqrt{{\partial^2 f_1\over{\partial u^2}} {\partial^2 
f_2\over{\partial v^2}}}}}
\eq
 where the sum is taken over  four different saddle points; 
 note that $f_1$, $f_2$ vanish  at these saddle points. 
We write down explicitly 
the expressions for $u(\lambda)$ and $v(\mu)$ by solving 
(\ref{5.34}) as
\begin{eqnarray}\label{5.36}
u(\lambda) &=& {1 - c^2\over{2}}\left ( - \lambda \pm 
\sqrt{ \lambda^2 - {4\over{1 - c^2}}}\right )
\nonumber\\
v(\mu) &=& {1 - c^2\over{2 c}} \left ( - \mu\pm 
\sqrt{ \mu^2 - {4 \over{
1 - c^2}}} \right )\nonumber\\
\end{eqnarray}
where we put $\lambda = \sqrt{{4\over{1 - c^2}}} {\rm sin} \theta$ 
and $ \mu =\sqrt{ {4 \over{ 1 - c^2}}} {\rm sin} \varphi$.
The saddle points become $u = i \sqrt{1 - c^2} e^{i \theta}, -i \sqrt{
1 - c^2}e^{- i\theta}$ and $v = {i\over{c}}\sqrt{1 - c^2}e^{i\varphi},
- {i\over{c}}\sqrt{1 - c^2} e^{- i\varphi}$.
Then adding these solutions of $u$ and $v$ in terms of $\theta$ 
and $\varphi$, we get from (\ref{5.35}), 
\begin{eqnarray}\label{5.37}
\rho^{II}(\lambda,\mu) &=&
{1 - c^2\over{8 N^2 \pi^2 c}} {1\over{{\rm cos}\theta {\rm cos}\varphi}}
\nonumber\\
&\times&\left ( 
{{2\over{c}} - ( 1 + {1\over{c^2}}) {\rm cos}( \theta - \varphi )
\over{[ ( 1+ {1\over{c^2}}) - {2\over{c}}{\rm cos} ( \theta - 
\varphi)]^2}} +
{{2\over{c}} + ( 1 + {1\over{c^2}}){\rm cos} ( \theta + \varphi )
\over{ [ ( 1 + {1\over{c^2}}) + {2\over{c}}{\rm cos} ( \theta + 
\varphi )]^2}}\right )\nonumber\\
\end{eqnarray}

This expression in the large N limit 
 coincides 
with the previous result \cite{BZ2}.
 The denominator
of (\ref{5.37}) does not vanish for $\lambda \rightarrow \mu$. 
Note that $c$ 
is related to the time $t$ as $c = e^{-t}$. When $t$ is small,
we have $ c \sim 1 - t$. Then, the denominator is 
approximated as $(1 - {1\over{c}})^2 + {(\theta - \varphi)^2
\over{c}} \sim t^2 + {t\over{2}}(\lambda - \mu )^2$, 
when $\lambda$ and $\mu$ are small. Note that we have to 
rescale $\lambda$ and $\mu$ for the time-dependent case 
by a factor $\sqrt{e^{-t}{\rm \sinh} t} = \sqrt{(1 - c^2)/2}$ 
as explained in (\ref{5.11}). Then we have $t^2 + {1\over{2}}
(\lambda - \mu)^2$ as a denominator and the result agrees 
with \cite{BZ2}.  In the time-dependent case in (\ref{5.1}), 
$\lambda$ and $\mu$ are interpreted as  one-dimensional 
space coordinates.

 In order to discuss the oscillatory  behavior, we return  to the expression  
(\ref{5.27}). We then  change 
  $z_2$ into $N z c$ and obtain 
\bq\label{5.38}
K_N(\lambda, \mu) = c N \oint {du\over{2\pi i}}
\int {dz\over{2\pi}} {1\over{u + i z}} e^{- N f( z, u )}
\eq
where $f( z u )$ is given by
\bq\label{5.39}
f( z,  u ) = {c^2 z^2\over{2 ( 1 - c^2)}} + i c \mu z - {\rm ln} z 
+ { u^2\over{2 ( 1 - c^2)}} + \lambda u + {\rm ln }u
\eq
Note that the variables $z$ and $u$ are decoupled, and the 
saddle point equations are simplified. 
Then, using the previous notaions $\lambda = \sqrt{{4\over{ 1 - c^2}}}
{\rm sin} \theta$ and $ \mu = \sqrt{{4\over{1 - c^2}}} {\rm sin}
\varphi$, we find the relevant 
saddle points for $z$ and $u$ from the solutions of  ${\partial 
f\over{\partial z}} = 0$ and ${\partial f \over{\partial u }} = 0$,
\begin{eqnarray}\label{5.40}
&&z = {\sqrt{1 - c^2}\over{c}} e^{- i \varphi}, {\hskip 5mm}
 - {\sqrt{1 - c^2}\over{c}} e^{ i \varphi}\nonumber\\
&& u = i \sqrt{1 - c^2} e^{i \theta}, {\hskip 5mm} 
 - i \sqrt{1 - c^2} e^{ - i \theta}   
\end{eqnarray}
For the saddle point values, $ z = {\sqrt{1 - c^2}\over{c}} e^{- i \varphi}$ and $ u = i \sqrt{ 1 - c^2} e^{i \theta }$, $f$ and the fluctuation 
determinant ${\partial^2 f\over{\partial u^2}} {\partial^2 
f\over{\partial z^2}}$ become
\begin{eqnarray}\label{5.41}
&&f ( z, u ) = i (\theta + \varphi ) +{1\over{2}} ( e^{2 i \theta} - 
e^{- 2 i \varphi} )
\nonumber\\
&& {1\over{\sqrt{{\partial^2 f\over{\partial u^2}}{\partial^2 f
\over{\partial z^2}}}}} = \left ( {1 - c^2\over{2 c}}\right ) { e^{ {i\over{2}}
(\theta - \varphi)}\over{\sqrt{{\rm cos}\theta {\rm cos }\varphi}}}
\end{eqnarray}

Adding other saddle points values, we obtain
\begin{eqnarray}\label{5.42}
&&K_N(\lambda, \mu ) = - { \sqrt{1 - c^2} \over{4 \pi N}}
{e^{{N\over{4}}( 1 - c^2) (\lambda^2 - \mu^2)}
  \over{\sqrt{ {\rm cos}\theta 
{\rm cos}\varphi} } }\nonumber\\
&\times&
 ( { {\rm cos} [ N ( h(\theta ) + h(\varphi))
- (\theta + \varphi)] + c {\rm cos} [ N ( h(\theta ) + h(\varphi))]
\over{ { 1 + c^2\over{2c}} + {\rm cos}
( \theta + \varphi ) }}
\nonumber\\
&-& {{\rm cos} [ N ( h(\theta ) - h(\varphi))
- (\theta - \varphi)] - c  {\rm cos} [ N ( h(\theta ) - h(\varphi))]
\over{ { 1 + c^2\over{2c}} - {\rm cos}
( \theta - \varphi ) }}  )\nonumber\\
\end{eqnarray}
where $h(\theta) = \theta + {1\over{2}}{\rm sin} 2\theta +
 {1\over{2N}}
\theta $.
Thus we find easily the oscillating behavior of $K_N(\lambda, \mu)$
from the integral representation in the large N limit. 
In \cite{Danna}, this result had been obtained 
through the integral representation of  Hermite polynomials.
Our derivation is more direct. 

The oscillating behavior of 
another kernel $\bar K_N(\lambda, \mu)$ is obtained similarly form
(\ref{5.27}).
If we make the smooth average for the product of $K_N(\lambda,\mu) $ and $\bar K_N(\lambda, \mu)$, we obtain the previous 
result (\ref{5.37}). Also, the expression of (\ref{5.42}) is the 
generalization of one matrix result of (\ref{a.4}), which may be 
obtained in the limit $c \rightarrow 1$.

The Fourier transform $K(\tau)$, the form factor is 
obtained f rom (\ref{5.42}). We consider the simple case 
$E = 0$.
\bq\label{5.43}
    S(\tau) = \int d\omega e^{i \omega \tau} \rho(E, E + \omega )|_{
E = 0}
\eq
Using (\ref{5.42}), we have for $\theta = 0, {2\over{\sqrt{1 - c^2}}}
{\rm sin} \varphi = \omega$,
\bq\label{5.44}
K_N(0,\omega) \bar K_N(0,\omega) = - {1 - c^2\over{
16\pi^2 N^2}} {1\over{{\rm cos}\varphi}}{1\over{[{(1 - c)^2\over{
4c^2}} + {\rm sin}^2 \varphi]^2}} f
\eq
where $f$ is
\bq\label{5.45}
  f = [ - 2 {\rm cos}\varphi{\rm cos}[N h(\varphi) - \varphi] + 
(1 + c^2 ){\rm cos} N h(\varphi)]^2
\eq
The Fourier transform of (\ref{5.43}) is an integral over $\omega$
which may be performed by taking the residue at the pole $ \omega = i{\sqrt{1 - c}\over{c}}$. Then we 
obtain 
\bq\label{5.46}
\int d \omega e^{i \omega \tau} K_N(0,\omega)\bar K_N(0, 
\omega) \simeq \tau e^{-{\sqrt{1 - c}\over{c}} \tau}
\eq
Thus we find that  the form factor has a linear term in $\tau$, but 
modified by  $e^{-{\sqrt{1 - c}\over{c}}\tau}$. Note that 
the integrand of (\ref{5.43})  has poles on both sides of the real axis
 the integral does not 
vanish for $\tau$  larger than $2a$  unlike (\ref{2.7}). However for large
 $\tau$, the form factor becomes exponentially 
small. The Fourier integral of the first term $\rho^{I}$ also 
becomes exponentially small for large $\tau$.
Therefore, the singularity of the form factor $S(\tau)$ at 
the Heisenberg time $\tau = 2N$ in one matrix model 
is smeared out for the time dependent case. 

 \sect{ Correlation functions with an external field}

   The  contour integral
method can be applied to the case  in which the  Hamiltonian is 
a sum of  a deterministic term and  of a random one 
\cite{BH}:
\bq\label{6.1}
     H = H_0 + V
\eq
$H_0$ is  a given deterministic term
 and $V$ is a random matrix with a  
 Gaussian distribution $P$  given by
\begin{eqnarray}\label{6.2}
P(H) &=& {1\over{Z}} e^{- {N\over{2}} {\rm Tr} V^2}
\nonumber\\
&=& {1\over{Z}} e^{-{N\over{2}} {\rm Tr} ( H^2 - 2 H_0 H + H_0^2)}
\end{eqnarray}
Then we are dealing with a Gaussian unitary ensemble modified 
by a matrix source $A = - H_0$.

Therefore, keeping the finite eigenvalues $a_i$ of $A$ and 
putting $H = M$, we readily obtain the 
correlation function in the presence of the deterministic term $A$.

By the contour-integral method we will show that 
 the connected part of the n-point 
correlation function is also given by the product of the two-
point kernel $K_N(\lambda,\mu)$
 in the case of a  nonvanishing external source
 $A$.

Let us recall the two-point correlation function, which becomes 
from (\ref{2.20}) \cite{BH},
\begin{eqnarray}\label{6.3}
U_A(t_1,t_2) &=& - {1\over{N^2}} \oint {dudv\over{(2\pi i)^2}}
e^{-{t_1^2\over{2N}} - {t_2^2\over{2 N}} - i t_1 u - i t_2 v}
 {1 \over{(u - v - { i t_1\over{N}})( u - v + {i t_2\over{N}})}
}\nonumber\\
&\times&\prod_{\gamma=1}^N ( 1 - {it_1\over{N(u - a_\gamma)}})( 1 - { it_2\over{
N(v - a_\gamma)}}) 
\end{eqnarray}
By the shift $t_1 \rightarrow t_1 - i u N$, we obtain the kernel
of the two point correlation function,
\bq\label{6.4}
K_N(\lambda, \mu) = {1\over{N}} \int {dt\over{2 \pi}}\oint 
{dv\over{2 \pi i}}\prod_{\gamma = 1}^{N}
( { a_\gamma + {i t\over{N}}\over{ v - a_\gamma}})
{1\over{v + {i t\over{N}}}} e^{- {N\over{2}}v^2 - { t^2\over{2N}} 
- i t \lambda - N v \mu}
\eq
The connected part of the two-level correlation function 
$\rho_c^{(2)}
(\lambda, \mu)$ is given by $- {1\over{N^2}}
K_N(\lambda,\mu)K_N(\mu,\lambda)
$.
The Fourier transform of the three-point correlation function 
is given by
\bq\label{6.5}
   U(t_1,t_2,t_3) = {1\over{N^3}}< {\rm Tr}e^{i t_1 M}
{\rm Tr}e^{i t_2 M} {\rm Tr} e^{i t_3 M}>
\eq
We evaluate this quantity by the same procedure for the 
two-point correlation function. 
We integrate out the eigenvalues of $M$ by the use of Itzykson-
Zuber formula. Then it becomes
\bq\label{6.6}
U(t_1,t_2,t_3) = {1\over{N^3}}\sum_{\alpha_i=1}^N
{\prod_{i<j}( b_i - b_j)\over{\prod_{i<j} ( a_i - a_j)}} e^{{N\over{2}} \sum 
b_i^2}
\eq
where $b_i$ is given
\bq\label{6.7}
b_i = a_i - {i t_1\over{N}}\delta_{i,\alpha_1}
 - {i t_2\over{N}}\delta_{i,\alpha_2} - {i t_3\over{
N}}\delta_{i,\alpha_3}
\eq 
We have to consider the cases; $\alpha_1=\alpha_2=\alpha_3$,
$\alpha_1=\alpha_2\ne\alpha_3$. These cases give the delta-
function part as (\ref{2.4}).

When all $\alpha_i$ are different, the contour integral representation is straightforward,
\begin{eqnarray}\label{6.8}
&&U(t_1,t_2,t_3) = \oint {du dv dw\over{( 2 \pi i)^3}}
{\prod(u - a_\gamma - { i t_1\over{N}})( v - a_\gamma
- {i t_2\over{N}})( w - a_\gamma - { i t_3\over{N}})
\over{\prod( u - a_\gamma)
(v - a_\gamma) ( w - a_\gamma)}}\nonumber\\
&\times& {( u - v - {it_1\over{N}} + { i t_2\over{N}}) ( u - w - { it_1\over{N}}
+ {it_3\over{N}})(v - w - { i t_2\over{N}}+ { i t_3\over{N}})
 \over{( u - v -{it_1\over{N}})(u - v + {it_2\over{N}})
( u - w - {it_1\over{N}})(u - w + {i t_3\over{N}})
(v - w - { it_2\over{N}})}}\nonumber\\
&\times&{(u - v)(v - w) (w - u) \over{( v - w + {i t_3\over{N}})}}
{1\over{i t_1 t_2 t_3}}
e^{- {t_1^2\over{2N}} - {t_2^2\over{2N}} - {t_3^2\over{2 N}} - i t_1 u 
- i t_2 v - i t_3 w}
\end{eqnarray}

Writing the part of the  numerators of (\ref{6.8}) as
\begin{eqnarray}\label{6.9}
&&[( u - v - {i t_1\over{N}})( u - v + {i t_2\over{N}}) - { t_1 t_2\over{
N^2}}] [ ( v - w - { i t_2\over{N}})( v - w + { i t_3\over{N}}) - { t_2t_3\over
{N^2}}]
\nonumber\\
&&\times [ ( u - w - {i t_1\over{N}})( u - w + {i t_3\over{N}})- {
t_1t_3\over{N^2}}]
\end{eqnarray}
 we have the following disconnected parts,
\bq\label{6.10}
U(t_1)U(t_2)U(t_3) - U(t_1)U(t_2,t_3)-U(t_2)U(t_1,t_3)
-U(t_3)U(t_1,t_2)
\eq
The remaining term is a connected term.
We consider the Fourier transform of this connected part $U_c(t_1,
t_2,t_3)$, and make the change of variables,
$t_1\rightarrow t_1 - i u N$,$t_2\rightarrow t_2 - i v N$
and $t_3 \rightarrow t_3 - i w N$. Then the remaining term of
(\ref{6.9}) becomes
simply
\bq\label{6.11}
(u + {it_2\over{N}})(v + {it_3\over{N}})(w + {it_1\over{N}})
+ (u + {it_3\over{N}})(v + {it_1\over{N}})(w + {it_2\over{N}})
\eq
After  cancellation of these terms with the corresponding 
terms in the denominator, we obtain the connected part of the 
three point correlation function $\rho^{(3)}(\lambda,\mu,\nu)$,
\bq\label{6.12}
\rho_c^{(3)}(\lambda,\mu,\nu) = K_N(\lambda,\mu)K_N(\mu,\nu)
K_N(\nu,\lambda) + K_N(\lambda,\nu)K_N(\nu,\mu)K_N(
\mu,\lambda)
\eq
where $K_N(\lambda,\mu)$ is given by (\ref{6.4}).
Thus we find that the three point correlation function has the 
same form as for the unitary Gaussian ensemble in terms of  
 the kernel $K_N(\lambda, \mu)$, but the kernel should be 
modified as (\ref{6.4}) for the external field case.

In the presence of the external source A, the kernel $K_N(\lambda, \mu)$
in (\ref{6.4}) is not expressed  as a sum of 
products of  orthogonal polynomials. However, it satisfies remarkably the 
following equation;
\bq\label{6.13}
\int_{-\infty}^{\infty}
 d\mu K_N(\lambda, \mu) K_N(\mu, \nu) = K_N(\lambda, \nu)
\eq
The proof of this equation is shown in Appendix B.

Although the kernel $K_N(\lambda, \mu)$ is no longer  
expressed in terms of  
 Hermite polynomials when the external source is present, it still posseses
 N simple eigenfunctions since for 
 $n \leq N - 1$,
\bq\label{6.14}
 \int_{-\infty}^{\infty} K_N(\lambda, \mu) H_n(\sqrt{N} \mu) e^{- {N\over{2}}
 \mu^2} d\mu
= H_n(\sqrt{N} \lambda ) e^{-{N\over{2}} \lambda^2}
\eq
where the $H_n(x)$ are the usual Hermite polynomials.
However for  $n\geq N$, (\ref{6.14}) does not hold. Indeed in the  case
 of  zero external source,
the right hand side of (\ref{6.14}) vanishes. When  the external
 source is non-zero,  the result is non-zero and is  $a_i$- dependent.
The proof of (\ref{6.14}) 
is given  in Appendix B.

The n-point correlation $R_n(\lambda_1,\cdots,\lambda_n)$ is defined 
\cite{Mehta} by
\bq\label{6.15}
    R_n(\lambda,\cdots, \lambda_n) = {N!\over{(N - n )!}} \int \cdots \int P_N
(\lambda_1, \cdots, \lambda_N) d\lambda_{n+1} \cdots d\lambda_N
\eq
\bq\label{6.16}
   P_N(\lambda_1,\cdot,\lambda_N) = <\prod_{i=1}^{N} {\rm Tr} 
\delta (\lambda_i - M)>
\eq
Without external source, this n-point correlation function 
is expressed  in terms of the kernel $K_N(\lambda_i,\lambda_j)$ as
\bq\label{6.17}
   R_n = {\rm det} ( K_N(\lambda_i, \lambda_j))
\eq
where $i,j = 1, \cdots, n$.
This result  was derived \cite{Mehta} by the use of (\ref{6.3}) since 
 one has
\bq\label{6.18}
      P_N(\lambda_1,\cdots,\lambda_n) = {1\over{N!}} {\rm det} (K_N(\lambda_i,
\lambda_j))
\eq
where $i,j = 1,\cdots,N$.
For a non-vanishing external source, from the representation (\ref{6.3})
, whose derivation is given 
 in Appendix B, it follows that  (\ref{6.17}) still holds. This result leads to the universality
 of the level spacing probability $P(s)$. As shown in a  previous work 
\cite{BH}, the kernel $K_N(\lambda, \mu)$ has a universal short 
distance behavior,
\bq
      K_N(\lambda, \mu) \simeq {{\rm sin}[ N \pi (\lambda - \mu) \rho 
({\lambda + \mu\over{2}})]\over{(\lambda - \mu) \pi}}
\eq
The level-spacing probability distribution $P(s)$ is related to the 
 probability of having an empty interval  with  width s  E(s),  by
\bq
     P(s) = {d^2\over{ds^2}} E(s)
\eq
and $E(s)$ is obtained from the integration of $P_N(\lambda_1,\cdots,\lambda_N)
$ in which the region $-{s\over{2}} < \lambda_i < {s\over{2}}$ is vacant.
Thus we write 
\bq
     E(s) = \prod_i (\int_{-\infty}^{\infty} - \int_{-{s\over{2}}}^{{s\over{2}}
}) d\lambda_i P_N(\lambda_1,\cdots,\lambda_N)
\eq
Since the kernel $K_N(\lambda,\mu)$ has universal form, same as the case 
without external source, we conclude that $P(s)$ becomes same as GUE.

\sect{Zeros of the kernel $K_N(\lambda, \mu)$}

Since at short distance the kernel $K_N(\lambda,\mu)$ exhibits oscillations,
and thus changes sign, this kernel must have 
 lines of zeros in the $(\lambda,\mu)$ 
plane. Note that the solution of the equation 
\bq\label{7.1}
  K_N(\lambda, \mu) = 0
\eq
are always  real in the case of the one matrix model. 
In the $(\lambda,\mu)$ plane the solutions of 
 (\ref{7.1}) lie on 2N lines as shown in Fig.2. 

When $\mu$ is large enough, $K_N(\lambda, \mu)$ is approximated 
by the product of  the Hermite polynomial $H_N(\lambda)$ multiplied by $\mu^N$ 
as shown in (\ref{A5}), and this Hermite 
polynomial has N real zeros.
These N real zeros give rise to N non-crossing lines in the $(\lambda,\mu)$
plane for finite $\mu$.
This behavior is a manifestation of the short distance universality 
in the large N limit. The distances between these  N lines become 
 equal when $\lambda - 
\mu$ is order of 1/N. 
These N lines are  parallel  to the line $\lambda = \mu$ when $\lambda$ 
and $\mu$ are inside  the support of the density of state.
At the edge, these lines bend, and  universality does not hold any more. 
  
We have discussed the time-dependent matrix model, which becomes 
equivalent to the two matrix model. In this two matrix model, the lines of  
zeros of $K_N(\lambda, \mu)$ show a  different behavior. In this case, the 
kernel $K_N(\lambda, \mu)$ is given by (\ref{5.31}) or (\ref{5.32}).
When $c \neq 1$, the lines of zeros of $K_N(\lambda, \mu)$ in the real 
$(\lambda,\mu)$ plane are not parallel to the  $\lambda = \mu$ line,  as 
shown in Fig.3 on which one can see that some lines turn over. Consequently
 along the line 
$\lambda = - \mu$, some  real solutions are missing. This behavior is 
related to the fact that the short-distance universality does not hold
in the two-matrix model. 

In the presence of an external source for the one matrix model, the 
solutions  of (\ref{7.1}) are not much modified by the source  
if its eigenvalues  are smoothly distributed.
This is also a  manifestation of the short distance universality for 
 the 
oscillatory behavior. 
When the ths support of the eigenvalues of the source are split
 into separate parts, this is reflected on the position of 
the lines of  zeros  
 as  shown in  Fig. 4.
The number of lines of zeros is  conserved for an  arbitrary distribution  of 
the external eigenvalues.

\sect{Discussion}

In this paper, we have  applied the contour-integral representation for the 
kernel which characterizes the  correlation functions, for the  one matrix 
model, the time-dependent matrix model and in the presence of an  external 
source.
We have investigated the form factor $S(\tau)$, which is the Fourier 
transform of the two-level correlation function, by the use of these 
contour-integral representations. The universality of the two-level 
correlation $\rho_c(\lambda, \mu)$ for $\lambda - \mu$ of 
order  1/N, implies immediately the linear behaviour of  $S(\tau)$ 
in the large N limit. We have found explicit  deviations from  the linear 
behavior of $S(\tau)$, and found  a new surprizing connection to 
the Laguerre ensemble for the average of $S(\tau)$.

Near the Heisenberg time $\tau = \tau_c$, a cross-over behavior 
is  observed. For the time- dependent matrix model, 
which we have mapped into an equivalent two -matrix model, the universal 
behavior of the one matrix model is no more present and
 the singularity at $\tau = \tau_c$ in $S(\tau)$ is then smeared out.
This behavior indicates that near the Heisenberg time, the form factor 
is not universal. The non-universality has been pointed out by the 
authors of \cite{Andreev1} for the case of  mesoscopic dirty 
metals. Our result is consistent with this non-universality.
Finally we have investigated the zeros of the kernel $K_N(\lambda, \mu)$
 for the
two matrix model, and found  differences  between the  one  and two matrix
 models. 

  For the matrix model with a non-zero external source, the universality of 
two-level correlation function holds, as shown in a  previous paper \cite{BH}.
With the technique of contour-integral representations  
 we have also  obtained all the higher correlation functions.

\vskip 5mm

{\bf Acknowlegment}

We are thankful for
  the support by the cooperative research 
project between the Centre National de la Recherche Scientifique 
and the Japan Society for the Promotion of Science.
S. H. thanks a Grant-in-Aid for Scientific Research by the Ministry of 
Education, Science and Culture.
\vskip 8mm

\setcounter{equation}{0}
\renewcommand{\theequation}{A.\arabic {equation}}
{\bf Appendix A: Integral representation for the kernel $K_N(\lambda,
\mu)$}
\vskip 5mm
It is known that the kernel $K_N(\lambda,\mu)$ is a 
sum of  products of  orthogonal polynomial $P_l(\lambda)$ and $P_l(\mu)$
\cite{Mehta}.
In the case of the Gaussian unitary ensemble, the orthogonal 
polynomials  $P_l(\lambda)$ are simply  Hermite polynomials.
We may  write these Hermite polynomials as  contour integrals
\begin{equation}\label{A1}
  H_l(\lambda) = \oint {du\over{2 \pi i}} {l!\over{u^{l + 1}}}
e^{\lambda u - {1\over{2}} u^2}
\end{equation}
Their  normalization  is  
\begin{equation}\label{A3}
 \int_{-\infty}^{\infty} d\lambda  H_l(\lambda)H_m(\lambda) 
e^{-{1\over{2}}\lambda^2}
=  {\sqrt{2 \pi}} l! \delta_{l,m}
\end{equation}
 It is convenient to use at the same time another integral
representation for these Hermite polynomials obtained 
by introducing an auxiliary variable  $t$,
\begin{eqnarray}\label{A4}
  H_l(\lambda) &=& \int_{-\infty}^{\infty} {dt\over{\sqrt{2\pi}}} 
\oint {du\over{2\pi i}} {l!\over{u^{l + 1}}} e^{\lambda u 
+ i t u - {1\over{2}} t^2}\nonumber\\
&=& \int_{-\infty}^{\infty} {dt\over{\sqrt{2 \pi}}} (i t)^l e^{
- {t^2\over{2 }} - i t \lambda + {\lambda^2\over{2}}}
\end{eqnarray}
The kernel $K_N(\lambda, \mu) $ is given by
\bq\label{A5}
  K_N(\lambda,\mu) = {\sqrt{{N\over{2 \pi}}}}
\sum_{l = 0}^{N - 1} {H_l(\sqrt{N}\lambda) 
H_l(\sqrt{N}\mu)\over{l!}} e^{- {N\over{2}} \lambda^2}
\eq
We use  two different expressions (\ref{A1}) and (\ref{A4}) 
of Hermite polynomials  in the kernel (\ref{A5}).
 The summation of the geometric series  give  $(1 - 
( {it\over{u}})^N)/1 - {it\over{u}}$.
Then we shift $t\rightarrow t/\sqrt{N} $ and $u\rightarrow - \sqrt{N}u$.

\bq\label{A6}
K_N(\lambda,\mu) = - \int_{-\infty}^{\infty}{ dt\over{2\pi}}
 \oint {du\over{2 \pi i}} (-{i t\over{N u}})^N
{1\over{ u + {i t\over{N}}}} e^{- {N\over{2}}u^2 - {1\over{2 N}}t^2 - i t 
\lambda - N u \mu}
\eq
This expression coincides with (\ref{2.22}), which has been derived 
earlier by Kazakov's method of a vanishing external source.

This integral representation may also be applied to the two-matrix model.
The expressions for $K_N(\lambda, \mu)$ and $\bar K_N(\lambda, \mu)$ 
in (\ref{5.27}) are obtained from (\ref{5.31}) and (\ref{5.32}) by 
the same integral representations of (\ref{A1}) and (\ref{A4}).

\vskip 8mm
 
\setcounter{equation}{0}
\renewcommand{\theequation}{B.\arabic {equation}}
{\bf Appendix B: The properties of the kernel $K_N(\lambda, \mu)$ in 
the external source}
\vskip 5mm
 We consider the proof of (\ref{6.13}). 
By the integral representation, we write the integrand, which is a product of
the kernels, by $I$, 
\begin{eqnarray}\label{B1}
I &=&  K_N(\lambda, \mu) K_N( \mu, \nu)\nonumber\\
 &=&  
\int_{\infty}^{\infty}
 {dt_1\over{2\pi}} \oint {du_1\over{2 \pi i}} \int_{\infty}^{\infty}
 {dt_2 \over{2 \pi}}
\oint {du_2\over{ 2 \pi i}} \prod_{\gamma} ({a_\gamma + {i t_1\over{N}}
\over{u_1 - a_\gamma}})({a_\gamma + {i t_2 \over{N}}\over{u_2 - a_\gamma}})
\nonumber\\
&\times& {1\over{u_1 + {it_1\over{N}}}} {1\over{u_2 + {i t_2\over{N}}}}
e^{ - {N\over{2}} u_1^2 - {N\over{2}} u_2^2 - {t_1^2\over{2 N}}
- {t_2^2\over{2 N}}  - i t_1 \lambda - i t_2 \mu - N u_1 \mu - N u_2 \nu}
\end{eqnarray}
Making the shift $t_2 \rightarrow t_2 + i N u_1$, and integrating $I$  over
 $\mu$, 
we obtain the $\delta(t_2)$ function. Thus the integral of I becomes
\begin{eqnarray}\label{B2}
\int_{-\infty}^{\infty}
 d\mu I &=&- (-1)^N \int_{-\infty}^{\infty} {dt_1\over{2\pi}} 
\oint {du_1\over{2\pi i}} \oint {du_2\over{2 \pi i}}
\prod_\gamma ({a_\gamma + {i t_1\over{N}}\over{u_2 - a_\gamma}}) {1\over{
(u_1 + {i t_1\over{N}})(u_2 - u_1)}}\nonumber\\
&\times& e^{ - {t_1^2\over{2 N}} - i t_1 \lambda 
- {N\over{2}} u_2^2 - N u_2 \nu}
\end{eqnarray}
The contour-integral of $u_1$ can be performed around $u_1 = - 
{i t_1\over{N}}$ since $u_1$ appears only in the denominator.
Then, we obtain 
\begin{eqnarray}\label{B3}
\int d\mu I &=& \int {dt_1\over{2 \pi}}\oint {du_2\over{2 \pi i}} 
\prod ( { a_\gamma + { i t_1\over{N}}\over{ u_2 - a_\gamma}}) {1\over{ 
u_2 + { i t_1\over{N}}}} e^{ - {N\over{2}} u_2^2 - {t_1^2\over{2 N}} 
- i t_1 \lambda - N u_2 \nu}\nonumber\\
&=& K_N(\lambda, \nu)
\end{eqnarray}

The integral equation of (\ref{6.14}) is verified similarly.
We evaluate first the following integral involving  Hermite polynomials,
\begin{eqnarray}\label{B4}
 \int_{-\infty}^{\infty} H_n(\sqrt{N} \mu) e^{-{N\over{2}} (\mu + u)^2} d\mu
  &=& \int_{-\infty}^{\infty} H_n(\mu) e^{-{1\over{2}}(\mu + \sqrt{N} u)^2}
{du\over{\sqrt{N}}}\nonumber\\
&=& \int_{-\infty}^{\infty} H_n(\mu) (\sum_{l=0}^{\infty} H_l( \mu)
{(-\sqrt{N}u)^l\over{l!}}) e^{-{1\over{2}}\mu^2}{d\mu\over{\sqrt{N}}}
\nonumber\\
&=& \sqrt{2\pi} (-u)^n N^{{n-1\over{2}}}
\end{eqnarray}
Using the expression of (\ref{6.4}) for $K_N(\lambda, \mu)$ with (\ref{B4}), 
we obtain 
\begin{eqnarray}\label{B5}
  &&\int K_N(\lambda, \mu) H_n(\sqrt{N} \mu) e^{-{N\over{2}} \mu^2} d\mu 
\nonumber\\
&=&
- (-1)^N
\sqrt{2\pi} N^{{n-1\over{2}}}
\int_{-\infty}^{\infty}{dt\over{2\pi}}\oint {du\over{2 \pi i}} 
\prod_{\gamma = 1}^{N} 
({a_\gamma + { i t\over{N}}\over{u - a_\gamma}}) {(-u)^{n}
\over{u + {i t\over{N}}
}} e^{- {t^2\over{2 N}} - i t \lambda}
\end{eqnarray}

When $n < N$, this contour-integration of $u$ converges for
 $|u|\rightarrow \infty$; we may then take the residues of the poles outside
of the contour , i.e. the pole  $u = - {it\over{N}}$
instead of the  $a_\gamma$'s.
Then, by evauating the residue, and performing the $t$ integration, we 
obtain
\begin{eqnarray}\label{B6}
  \int_{-\infty}^{\infty} K_N(\lambda, \mu) H_n(\sqrt{N} \mu) e^{- {N\over{2}}
\mu^2} d\mu &=& \int_{-\infty}^{\infty} {dt\over{\sqrt{2\pi}}} ({it\over{
N}})^n N^{{n-1\over{2}}}e^{-{1\over{2N}}t^2 - i t \lambda}\nonumber\\
&=& H_n(\sqrt{N}\lambda) e^{-{N\over{2}}\lambda^2}
\end{eqnarray}
Note that the contour-integration around infinity does not 
converges  for $n\geq N$; we cannot take the poles outside of the contour
any more and  we obtain a  different result.

The kernel $K_N(\lambda, \mu)$ for  a non-zero external source can be written 
as  a determinant; this is useful for numerical calculations.
Since the kernel $K_N(\lambda, \mu)$ satisfies (\ref{B6}), we write 
an expression for $K_N(\lambda, \mu)$ as a determinant, in which 
the variable $\mu$ appears only in the first row of the $N\times N$ matrix.
The eigenvalues of external source $a_i$ appears only in the i-th column 
of the matrix. This is related to the fact that the exchange between $a_i$ 
and $a_j$ in (\ref{B7}) does not affect  $K_N(\lambda,\mu)$.
We can write the matrix element of the i-th row as a polynomial of 
order of (i - 1) for $\lambda$.
 The first three rows, for example,  are expressed by
\begin{eqnarray}\label{B7}
 K_N(\lambda, \mu) &=& - {1\over{\sqrt{2\pi}}
}{e^{- {N\over{2}} \lambda^2}\over{\prod_{i<j} (a_i - a_j)}}\nonumber\\
&\times&
\det\left(\matrix{ e^{-{N\over{2}}a_1^2 - N a_1 \mu}& 
e^{-{N\over{2}} a_2^2 - N a_2 \mu}&\ldots&
 e^{ -{N\over{2}} a_N^2 - N a_N \mu}\cr
\lambda + a_1& \lambda+ a_2&\ldots&\lambda + a_N\cr
(\lambda+a_1)^2 + {1\over{N}}& (\lambda + a_2)^2 + {1\over{N}}& 
\ldots&(\lambda + a_N)^2 + {1\over{N}}\cr
& \ldots&\ldots &}
\right)\nonumber\\
\end{eqnarray}
The matrix element $m_{ij}$ is $(\lambda + a_j)^{i - 1}$ when i is even.
When i is odd, then $m_{ij}$ is given by $(\lambda + a_j)^{i -1} + C$,
where C is a constant depending  upon N. This constant is determined to be 
consistent with (\ref{B6}).  

As a simple example for N=2, we obtain
\bq\label{B8}
   K_2(\lambda,\mu) = - {1\over{\sqrt{2\pi}}}
{1\over{(a_1 - a_2)}} e^{-\lambda^2}
     \det \left ( \matrix{e^{- a_1^2 - 2 a_1 \mu}& e^{-a_2^2 - 2 a_2 \mu}\cr
       \lambda + a_1 & \lambda + a_2}\right )
\eq
It may be interesting to note that there  exists values of the $a_i$ for
which  the two level correlation function
$\rho_c(\lambda,\mu) = - K_N(\lambda,\mu)K_N(\mu,\lambda)$ ,
 which is normally negative,
may become positive. 
For instance, in the case $N=2$, we consider  $\mu = 0$,
 $a_2 = 0$. In this case, we obtain from (\ref{B8}),
$K_2(\lambda,0) = -1/\sqrt{2\pi}
(\lambda e^{-a_1^2} - \lambda - a_1)e^{-\lambda^2}/a_1$ 
and $K_2(0,\lambda) = 1/\sqrt{2\pi}$. Then, we have
\bq
      \rho_c(\lambda, 0) = {1\over{8\pi a_1}}( \lambda e^{-a_1^2} - \lambda - a_1) e^{-\lambda^2}
\eq
This expression may take positive values, for example when $a_1 = 1$ and 
$\lambda < -1.6$.
The expression (\ref{B7}) is useful for locating numerically the
zeros of $K_N(\lambda, \mu)$.

Finally let us mention that the contour-integral technique which we have used here may be extended to 
 the n-point functions. If we define

\begin{eqnarray}
   U(t_1,\cdots,t_n) &=& \oint {\prod du_i\over{(2\pi i)^n}}
\prod_{i=1}^{n}\prod_{\gamma=1}^{N} {(u_i - a_\gamma - {i t_i\over{N}})\over
{( u_i - a_\gamma)}} \prod_{i<j} {(u_i - u_j - { it_i\over{N}} + {i t_j\over{
N}})\over{( u_i - u_j - {i t_i\over{N}})(u_i - u_j + {i t_j\over{N}})}}
\nonumber\\
&\times& \prod_{i<j} ( u_i - u_j) {1\over{\prod t_i}} e^{ - {1\over{2 N}}
\sum t_i^2 - i \sum t_i u_i}
\end{eqnarray}
 then $R_n(\lambda_1,\cdots.\lambda_n)$ is obtained by  Fourier 
transform of $U(t_1,\cdots,t_n)$ as
\bq
   R_n(\lambda_1,\cdots,\lambda_n) = \int U(t_1,\cdots,t_n) e^{- i \sum t_i \lambda_i} dt_i
\eq

\newpage


\newpage
\vskip 5mm
{\bf Figure caption}
\begin{itemize}
\item Fig. 1a,  A quasi-linear behavior of $S(\tau)$,
N = 16
\item Fig. 1b, The derivative of $S(\tau)$ of Fig. 1a.
\item Fig. 2.
The zeros of $K_N(\lambda, \mu)$ are plotted as lines in the real 
$(\lambda, \mu)$- plane, for N = 5.
\item Fig. 3. The lines of zeros of $K_N(\lambda, \mu)$, N = 5 for the two-matrix
model with $c = {1\over{2}}$.
\item Fig. 4. The lines of zeros of $K_N(\lambda, \mu)$, N = 5 for the external 
source, $a_1 = -2, a_2 = 2, a_3 = 2.25, a_4 = 2.5$ and $a_5 = 2.75$
\end{itemize}

\vskip 5mm
\begin{thebibliography}{99} 
\bibitem{Mehta}  M. L. Mehta, Random matrices, 2nd ed. 
                   (Academic Press, New York 1991).
\bibitem{Dyson} F. Dyson, Journ. Math. Phys.{\bf 13}, (1972) 90.
\bibitem{Berry} M. V. Berry, Proc. Roy. Soc. A{\bf 400}, 229 (1985).
\bibitem{Andreev1} A. V. Andreev and B. L. Altshuler,
               Phys. Rev. Lett.{\bf 75} (1995) 902.
\bibitem{AgamAltAnd} O. Agam, B. L. Altshuler and A. V. Andreev,
                  Phys. Rev. Lett. {\bf 75} 4389 (1995).
\bibitem{AgamFish} O. Agam and S. Fishman, Phys. Rev. Lett. {\bf 
               76} (1996) 726.
\bibitem{BH} E. Br\'ezin and S. Hikami, to be published in Nucl. Phys. B
.( LPTENS-96-23/
             UT-Komaba-96-11. cond-mat/9605046).
\bibitem{Kazakov} V. A. Kazakov, Nucl. Phys. {\bf B 354} (1991) 614.
\bibitem{BHZ1} E. Br\'ezin, S. Hikami and A. Zee, Nucl. Phys. 
           {\bf B 464} (1996), 411.
\bibitem{HZ}   S. Hikami and A. Zee, Nucl. Phys. {\bf B 446} (1995) 337.
\bibitem{NS} K. Slevin and T. Nagao, Phys. Rev. Lett. {\bf 70} 
                     635, (1993).
\bibitem{Andreev} A. V. Andreev, B. D. Simons and N. Taniguchi,
                Nucl. Phys. {\bf B432}, 487 (1994) 487.
\bibitem{BHZ2} E. Br\'ezin, S. Hikami and A. Zee, Phys. Rev {\bf E 51}
                (1995) 5442.
\bibitem{Itzykson} C. Itzykson and J. -B. Zuber, J. Math. Phys. 
                {\bf 21} (1980) 411.
\bibitem{BZ2} E. Br\'ezin and A. Zee, Phys. Rev. {\bf E 49} (1994) 2588.
\bibitem{Beenakker} C. W. J. Beenakker, Phys. Rev. Lett. {\bf 70}
               1155 (1993).
\bibitem{Szafer} A. Szafer and B. L. Altschuler, Phys. Rev. 
               Lett. {\bf 70} 587 (1993).
\bibitem{Shastry}  O. Narayan and B. S. Shastry, Phys. Rev. Letts. 
                {\bf 71}, 2106 (1993).
\bibitem{Danna} D'anna, E. Br\'ezin and A. Zee, Nucl. Phys. {\bf 443}
              (1995) 433.
\bibitem{BZ1} E. Br\'ezin and A. Zee, Nucl. Phys. {\bf B 402} 
(1993) 613.
\end{thebibliography}
\end{document}